\documentclass[prb,10pt,twocolumn,superscriptaddress,aps,amsmath,amsfonts,notitlepage,longbibliography]{revtex4-2}
\usepackage{amsmath,amsfonts,amssymb,dsfont,graphicx,bm}
\graphicspath{{Figures/}}
\usepackage{color}
\usepackage{hyperref}
\usepackage{tikz}
\usepackage{pbox}
\usepackage{balance}

\usepackage{cancel}

\usepackage{txfonts}

\usepackage[normalem]{ulem}

\def\be{\begin{equation}}
\def\ee{\end{equation}}
\def\bea{\begin{eqnarray}}
\def\eea{\end{eqnarray}}
\def\bpm{\begin{pmatrix}}
\def\epm{\end{pmatrix}}

\newcommand{\Eqref}[1]{Eq.~\eqref{#1}}

\def\Re{\mathop{\rm Re}}

\def\diag{\mathop{\rm diag}}

\def\diff{\mathrm{d}}

\newcommand{\p}{\partial}

\begin{document}
\title{GHz non-reciprocal optical conductivity in hematite ($\alpha$-$\text{Fe}_2\text{O}_3$)}

\author{Peng Rao}
\thanks{These authors contributed equally to this work.}
\affiliation{Technical University of Munich, TUM School of Natural Sciences, Physics Department, 85748 Garching, Germany}

\author{Johannes Gr\"obmeyer}
\thanks{These authors contributed equally to this work.}
\affiliation{Technical University of Munich, TUM School of Natural Sciences, Physics Department, 85748 Garching, Germany}
\author{ P. Peter Stavropoulos}
\affiliation{Institut f\"ur Theoretische Physik, Goethe-Universit\"at Frankfurt, 60438 Frankfurt am Main, Germany}

\author{Alexander Mook}
\affiliation{Institut für Festkörpertheorie, Wilhelm-Klemm-Straße 10, 48149 Münster, Germany}

\author{Matthias Althammer}
\affiliation{Technical University of Munich, TUM School of Natural Sciences, Physics Department, 85748 Garching, Germany}
\affiliation{Walther-Mei{\ss}ner-Institut, Bayerische Akademie der Wissenschaften, 85748 Garching, Germany}

\author{Hans Huebl}
\affiliation{Technical University of Munich, TUM School of Natural Sciences, Physics Department, 85748 Garching, Germany}
\affiliation{Walther-Mei{\ss}ner-Institut, Bayerische Akademie der Wissenschaften, 85748 Garching, Germany}
\affiliation{Munich Center for Quantum Science and Technology (MCQST), Schellingstr. 4, 80799 München, Germany}

\author{Alexander Holleitner}
\affiliation{Technical University of Munich, TUM School of Natural Sciences, Physics Department, 85748 Garching, Germany}
\affiliation{Munich Center for Quantum Science and Technology (MCQST), Schellingstr. 4, 80799 München, Germany}
\affiliation{Walter Schottky Institute, Am Coulombwall 4a, 85748 Garching, Germany}

\author{Johannes Knolle}
\affiliation{Technical University of Munich, TUM School of Natural Sciences, Physics Department, 85748 Garching, Germany}
\affiliation{Munich Center for Quantum Science and Technology (MCQST), Schellingstr. 4, 80799 München, Germany}

\date{\today}
\begin{abstract}
We study the non-reciprocal properties of the iron oxide $\alpha$-Fe$_2$O$_3$ (hematite) in the canted easy-plane antiferromagnetic phase, specifically in the GHz to THz frequency range. First, using the the microscopic spin Hamiltonian, we obtain the correct classical ground state where the canting is induced by the Dzyaloshinskii-Moriya interactions (DMI). The magnon spectrum is simulated using linear spin wave theory. We then compute the polarizability and the sub-gap optical conductivities using linear response. We find that the conductivity tensor contains frequency peaks at the zero momentum magnon gaps of order $0.1~$meV which can be tuned by the DMI and on-site anisotropic spin interactions. Furthermore, we show that the canting-induced net magnetic moment $\mathbf{m}$ represents a measure for the effective time-reversal-symmetry breaking and non-reciprocity of the system: a finite $\mathbf{m}$ results in a non-zero Hall conductivity. Finally, we discuss the prospective application of hematite in non-reciprocal circulator design, by computing the non-reciprocal circulator transmission amplitude using the conductivities as input. 

\end{abstract}

\maketitle

\section{Introduction}

Systems with broken time-reversal-symmetry (TRS) can exhibit initially unexpected transport properties, such as the quantum anomalous Hall (QAH) effect~\cite{chang2023colloquium} where non-dissipative electric currents flow upon application of an external voltage. Typically, QAH arises from non-trivial band topology of electrons. In a topological insulator (TI), the non-trivial topological bands are gapped which, on a finite sample, gives rise to topological edge modes whose number is equal to the sum of band Chern numbers below the chemical potential. The edge modes carry charge and the off-diagonal QAH conductivity is quantized as a result. This has found important applications in recent years in manufacturing non-reciprocal devices such as circulators~\cite{mahoney2017chip,mahoney2017zero} that require large off-diagonal Hall conductivities. Usually, TRS breaking is achieved by an external magnetic field which however makes miniaturization of devices difficult. Alternatively TRS breaking can occur intrinsically. This is the case for ferromagnetic topological thin films, such as  Cr-doped Bi$_2$Te$_3$ or Sb$_2$Te$_3$~\cite{chang2023colloquium}. However, in these materials the Curie temperatures are of the order of a few Kelvin which make their implementation in practical circulators difficult.

Intrinsic TRS breaking also occurs in a much more common setting, i.e. magnetic insulators whose critical temperatures can be comparable or higher than room temperature. At first sight, their properties cannot be related to the QAH effect because the system is strongly insulating and the electric charges are localized near each magnetic atom. Indeed, their DC conductivities are identically zero. However, optical absorption and a non-zero AC conductivity have been described for electromagnetic excitations below the charge gap of insulators with magnetic moments~\cite{ng2007power,tokura2014multiferroics,bolens2018mechanism}. Microscopically, such an AC conductivity can be understood in terms of dynamic charge fluctuations as the time-varying electric field can polarize the insulator periodically~\cite{bolens2018mechanism}. Following this conjecture, the question  arises  whether the QAH effect can be also present in magnetic insulators with broken TRS. We emphasize that the microscopic mechanism for a finite AC conductivity in magnetic insulators is qualitatively different to the one in a TI, where DC charge transfer is due to the topological edge modes. In this work, we demonstrate that an AC component of the QAH can arise from the aforementioned below-gap mechanism also in magnetic insulators~\cite{ng2007power,tokura2014multiferroics,bolens2018mechanism} and we exemplify the developed theoretical framework in the case of $\alpha$-Fe$_2$O$_3$ (hematite).

Hematite is a canted easy-plane antiferromagnet with a finite net magnetization at room temperatures (referred to as the weak ferromagnetic phase). The system transitions into an easy-axis antiferromagnet below the Morin temperature $T_\text{M}\approx 260~$K~\cite{morin1950hematite} and becomes a trivial paramagnet above the Ne\'el temperature $T_\text{N}\approx 950~$K. Despite the magnetic moments being almost anti-parallel, they are canted due to the relativistic Dzyaloshinskii-Moriya interactions (DMI) to give a non-zero net magnetic moment $\mathbf{m}$. We find that a finite $\mathbf{m}$ breaks the effective TRS of the AFM to give a non-zero QAH effect, and the persistence of weak ferromagnetism into high temperatures makes hematite potentially useful for the QAH effect at ambient conditions. Moreover, the DMI energy scale, which relates to TRS breaking, is of the order of $0.1~$meV~\cite{dannegger2023magnetic}. This is in the GHz frequency range at which circulators operate.

In this paper, we consider the conductivity of the hematite in the weak ferromagnetic phase and the application to circulators. In Section~\ref{sec:model}, we analyze the spin Hamiltonian using linear spin wave theory (LSWT) to find the classical ground state and the magnon spectrum. At low frequencies and small momenta, two gapped magnon branches are present, corresponding to in-plane and out-of-plane oscillations of the magnetic moments respectively. The in-plane magnons are gapped by the triaxial basal anisotropy whereas the out-of-plane magnon gap is due to the DMI and the second order on-site anisotropy. We also discuss the dependence of the low-energy magnon gaps on material parameters. In Section~\ref{sec:non-reciprocal}, we show that the conductivity receives contributions from both magnon branches when the frequency is at the magnon gap. In particular, the system exhibits a QAH conductivity which receives contributions only from the out-of-plane oscillations. In Sec.\,\ref{sec:circulators}, we consider the potential application of the off-diagonal conductivities in hematite in a capacitive 3-port circulator geometry, similar to Refs.~\cite{mahoney2017chip, mahoney2017zero}, with a capacitive coupling between the high-frequency circulator ports and hematite.

\begin{figure}
    \centering
    \includegraphics[width=1\linewidth]{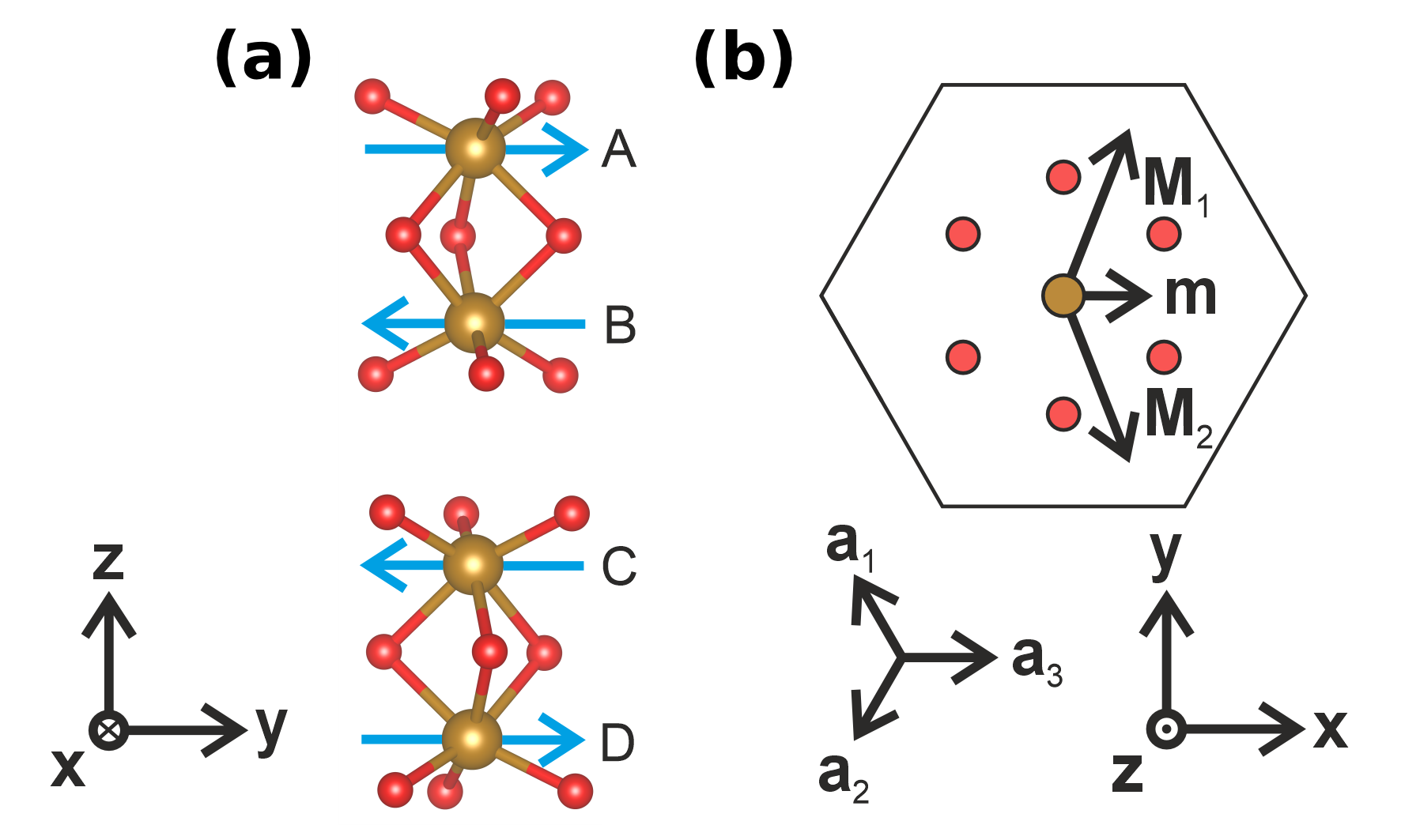}
      \caption{Illustration of the unit cell of hematite ($\alpha-\mathrm{Fe}_2\mathrm{O}_3$), showing iron atoms in brown and oxygen atoms in red. (a) Side-view of the unit cell with the magnetic moments shown in blue. (b) Top view looking down the $c$-axis. The in-plane projection of the lattice and the vectors $\mathbf{a}_{1},\mathbf{a}_{2}$ and $\mathbf{a}_{3}$ are shown. The magnetic moments $\mathbf{M}_1$ (sublattices $A,D$) and $\mathbf{M}_2$ (sublattices $B,C$) are almost anti-parallel and in-plane.}
    \label{fig:schematic}
\end{figure}

\section{Spin Hamiltonian and Linear spin wave theory}\label{sec:model}

Hematite belongs to the symmetry group $R\bar{3}c$ with four Fe$^{3+}$ magnetic sublattices stacked along the perpendicular $c$-axis, see Fig.\,\ref{fig:schematic}. In this paper, we use the coordinate system introduced in Ref.~\cite{hoyer2025altermagnetic}, where the $c$-axis is parallel to the $z$-axis.

To model this system, we use the spin Hamiltonian from Ref.~\cite{dannegger2023magnetic} which sums the contributing spin-spin interactions and the crystalline anisotropy
\begin{equation}\label{eq:spin-Hamiltonian}
H = H_{\text{exch}} + H_{\text{DMI}} + H_{\text{ani}}.
\end{equation}
The first term $H_{\text{exch}}$ describes the Heisenberg exchange interaction between the spins:
\begin{equation}
   H_{\text{exch}} =  \sum_{i,j \in n} J_n \mathbf{S}_i.\mathbf{S}_j ,
\end{equation}
where the summation is over the $n$-th nearest neighbor (N.N.) with the corresponding exchange constants $J_n$. Following Ref.~\cite{dannegger2023magnetic} we include up to the fifth nearest neighbor exchange couplings. Unless stated otherwise, their values are as follows (in units of meV):
\begin{align}\label{eq:spin-Hamiltonian-exchange}
    &J_1 = 3.21,~J_2 =3.84,~J_3=26.10,~J_4=15.71,J_5=0.17.
\end{align}
We neglect interactions originating from spins with further distances as we focus on low-energy magnons. A magnon branch splitting at higher energies, which is associated with altermagnetism is discussed by Hoyer et al.\,\cite{hoyer2025altermagnetic}, by accounting for interactions up to the $13$th nearest neighbor. However, the energies at which this effect become pronounced are far higher than the low energy range considered in this work, therefore we do not include $J_{13}$.
 
The second term $H_{\text{DMI}}$ contains the DMI between spins connected by $n$-th N.N. bonds:
\begin{equation}\label{eq:spin-Hamiltonian-DMI}
     H_{\text{DMI}}  = \sum_{i,j\in n}\mathbf{D}^{(n)}_{ij}.(\mathbf{S}_i \times \mathbf{S}_j) 
\end{equation}
We include the DMI up to the fourth N.N.; due to the presence of local inversion centers at next N.N. bonds, $\mathbf{D}^{(2)}$ are zero. The values and directionality of $\mathbf{D}^{(n)}$ follow that of Ref.~\cite{dannegger2023magnetic}; see Fig.~4 there. The non-zero DM vectors $\mathbf{D}^{(n)}$ are (in units of meV):
\begin{equation}
  D^{(3)}_x= D^{(3)}_z = 0.18; \ D^{(4)}_x = -0.09, \  D^{(4)}_y = -0.18, \  D^{(4)}_z = -0.15. 
\end{equation}
Note that the $\mathbf{D}^{(1)}$ vectors are small and we neglect them here.

The last term in Eq.~\eqref{eq:spin-Hamiltonian} are on-site anisotropies induced by the crystal lattice:
\begin{equation}\label{eq:spin-Hamiltonian-anisotropy}
    H_{\text{ani}} = d_2 \sum_i (S^z_i)^2 + d_6\sum_i \left[(S^+_i)^3 + (S^-_i)^3\right]^2.
\end{equation}
The first term in Eq.~\eqref{eq:spin-Hamiltonian-anisotropy} is the second order on-site anisotropy. As we discuss in Sec.~\ref{sec:GS}, a positive $d_2$ stabilizes the in-plane canted AFM configuration. For negative $d_2$ the classical spin configuration is the easy-axis AFM along the $z$-axis. Since we are interested in temperatures above $T_\text{M}$, we shall take:
\begin{equation}
    d_2 = 10^{-4}~\text{meV}.
\end{equation}
The second term in Eq.~\eqref{eq:spin-Hamiltonian-anisotropy} is a triaxial basal anisotropy allowed by symmetry. Such an anisotropy of order $1$~neV has been reported experimentally~\cite{banerjee1963attempt,flanders1964anisotropy}; on the theoretically level, the triaxial basal anisotropy is generated by order-by-disorder corrections which give the anisotropy energy scale $10^{-8}$~meV~\cite{hoyer2025altermagnetic}. The $d_6$ term takes this effect into account phenomenologically. Its role will be explained in Secs.~\ref{sec:GS} and \ref{sec:pGgap}. For now, we state the value of $d_6$:
\begin{equation}\label{eq:model-d6}
    d_6 = 10^{-8}~\text{meV}.
\end{equation}
As we shall see in Sec.~\ref{sec:pGgap}, the pseudo-Goldstone gap is extremely sensitive to $d_6$ and that even such a small $d_6$ can already induce a non-negligible magnon gap. This is the exchange-enhancement effect known in AFM materials, where the magnon gap scales as $\Delta \sim \sqrt{J_3 d_6}$ where $J_3$ is the largest exchange energy in \eqref{eq:spin-Hamiltonian-exchange}. 
Note also that $d_6$ comes with the large spin factor $S^6 =(5/2)^6 \approx 244$, and that the hierarchy of energy scales is $J_n \gg |\mathbf{D}^{(n)}| \gg  d_2\gg d_6$.

To analyze the Hamiltonian~\Eqref{eq:spin-Hamiltonian} using linear spin wave theory, we develop a Mathematica notebook that automatically finds the classical ground state, rotates the spins into the local polarized axis and performs the Holstein-Primakoff (HP) approximation~\cite{code}; the analytical details of the implementation can be found in Appendix~\ref{sec:LSW}. In the rest of this section, we discuss the qualitative features of each step and then, present the magnon spectrum.


\subsection{Classical ground state}\label{sec:GS}

The classical ground state (GS) spin configuration is a canted antiferromagnet having almost anti-parallel moments $\mathbf{M}_1$ (sublattices $A,D$) and $\mathbf{M}_2$ (sublattices $B,C$). The small $d_2$ restricts $\mathbf{M}_1$ and $\mathbf{M}_2$ to the $xy$-plane (see Fig.~\ref{fig:schematic}), where in the absence of $d_6$, the ground states are degenerate with respect to in-plane directions~\cite{hoyer2025altermagnetic}. This can be seen from the DMI term \Eqref{eq:spin-Hamiltonian-DMI} where for a given $n$-th N.N., the summation over $\mathbf{D}_{ij}^{(n)}$ is taken over at least three bonds related by C$_3$ symmetry. In the classical limit, the in-plane components of $\mathbf{D}_{ij}^{(n)}$ sums to zero and the system has planar rotation symmetry. The DMI, however, cants $\mathbf{M}_1$ and $\mathbf{M}_2$ and induces a small net magnetic moment $\mathbf{m} = \mathbf{M}_1+ \mathbf{M}_2$. 
We find the canting angle to be $\kappa \approx 0.058^\circ$ which is close to the experimental result $\kappa \approx 0.055^\circ$~\cite{hill2008neutron}. The classical in-plane degeneracy is removed by finite $d_6$. Classically, the basal anisotropy can be written as
\begin{equation}
       V_6 \sim 2 d_6 S^6 \cos 6\theta + \text{const.} ,
\end{equation}
where $\theta$ is the in-plane polar angle of one of the spin vectors. Thus, the ground state is in-plane and sixth-fold degenerate. In what follows, we choose the ground state with $\mathbf{M}_{1,2}$ approximately along the $y$-axis ($\theta \approx \pm \pi/2$) such that $\mathbf{m} \parallel \mathbf{e}_x$. As will be shown in Sec.~\ref{sec:pGgap}, the associated pseudo-Goldstone mode when $d_6=0$ corresponds to in-plane oscillations and acquires a gap from a finite $d_6$. This mode contributes to the Hall conductivities. 


\subsection{The magnon spectrum}\label{sec:pGgap}

\begin{figure}
    \centering
    \includegraphics[width=1\linewidth]{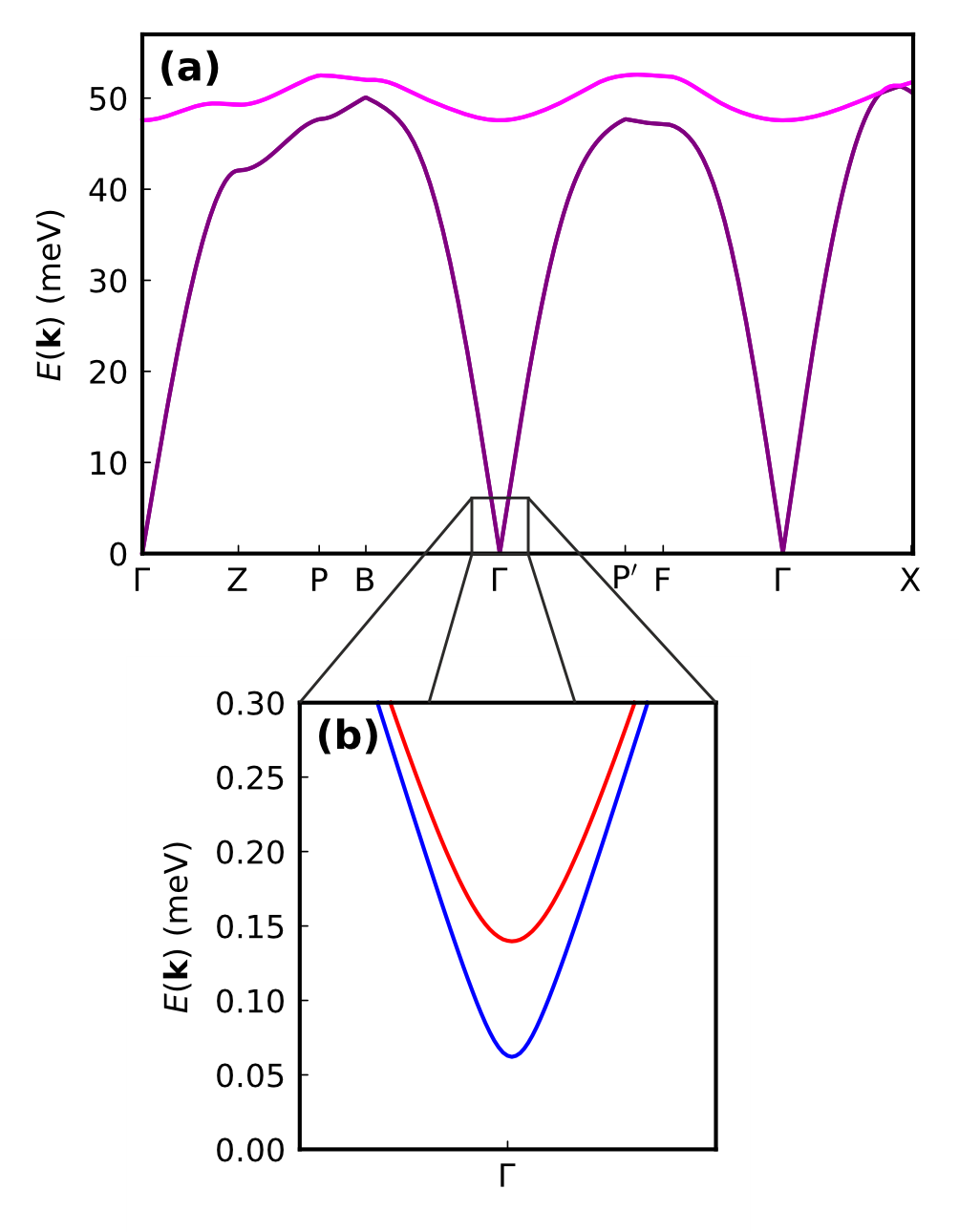}
    \caption{Linear spin wave spectrum for the spin Hamiltonian~\Eqref{eq:spin-Hamiltonian}. (a) Magnon spectrum across the Brillouin zone (BZ) of $\alpha$Fe$_2$O$_3$. Here, four magnon branches are pair-wise almost degenerate (magenta and purple lines for each pair). (b) The magnon spectrum near the $\Gamma$-point. The lower magnon branch corresponds to in-plane oscillations of the Ne\'el vector and the higher branch to out-of-plane oscillations.}
    \label{fig:LSW}
\end{figure}

We numerically find the linear spin wave spectrum of the spin Hamiltonian~\Eqref{eq:spin-Hamiltonian} with the aforementioned ground state. In Fig.~\ref{fig:LSW}(a), we see that the four magnon branches are almost pair-wise degenerate across the Brillouin zone (BZ). There are two magnon branches at low-energies near $\mathbf{k}=0$ ($\Gamma$ point) with gaps $0.06$~meV and $0.15$~meV respectively; see Fig.~\ref{fig:LSW}(b). The lower branch (blue line) corresponds to in-plane oscillations that are gapped by the triaxial basal anisotropy $d_6$. The higher branch (red line) corresponds to the out-of-plane oscillations and is gapped out by the DM interactions and the second order on-site anisotropy. In particular, the gap is comparable with the DMI scale We verify that with only $H_{\text{exch}}$, both branches become gapless.  

\begin{figure}
    \centering
    \includegraphics[width=1\linewidth]{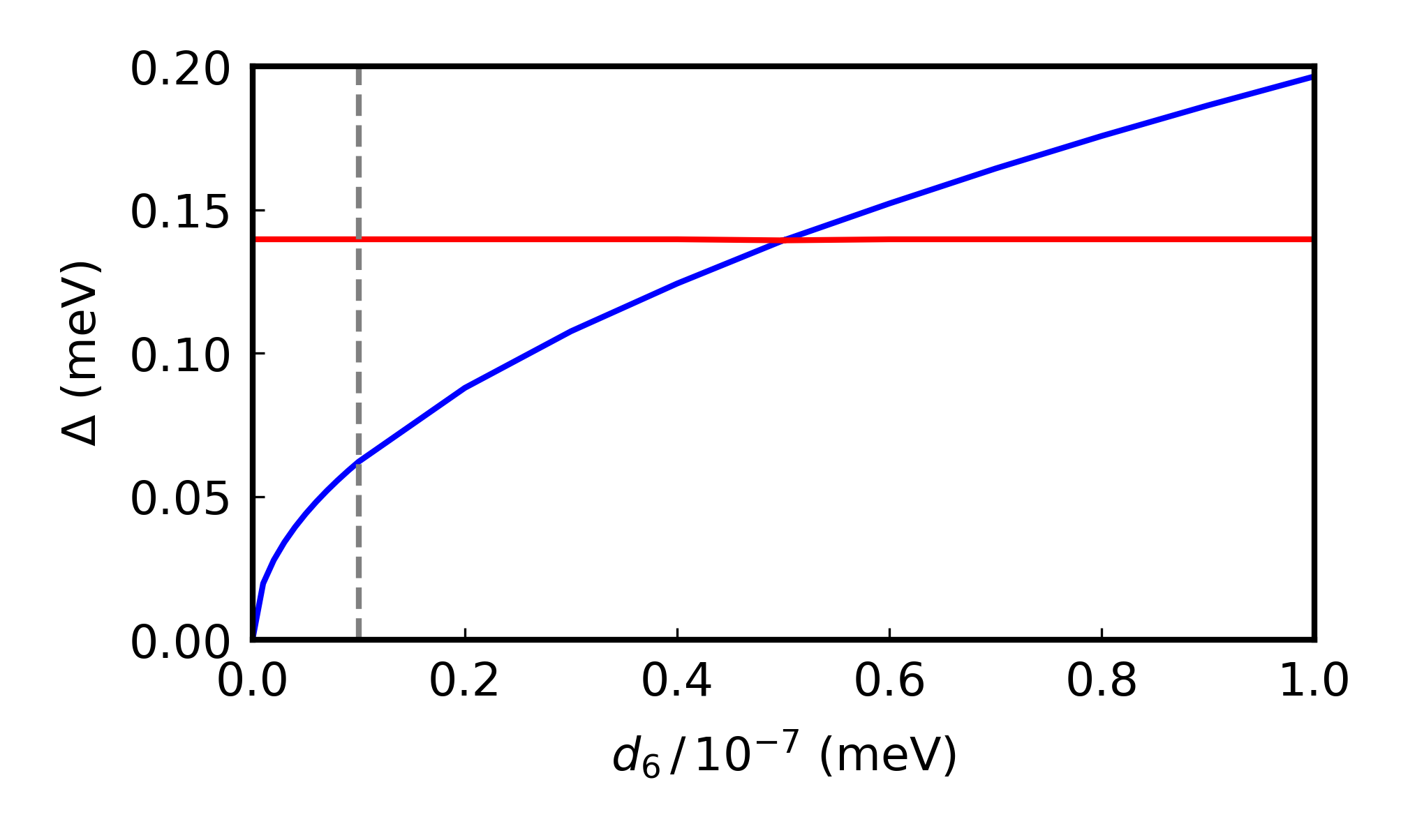}
    \caption{Minimal magnon gaps energy of the two lower branches in Fig.~\ref{fig:LSW}(b) as a function of the triaxial basal anisotropy strength $d_6$ in \Eqref{eq:spin-Hamiltonian-anisotropy}. The gray dashed line corresponds to $d_6=10^{-8}~$meV used elsewhere in the paper. Only the blue branch depends on $d_6$ since it corresponds to in-plane oscillations. The strong dependence on $d_6$ is due to the exchange enhancement effect discussed below Eq.~\eqref{eq:model-d6}. }
    \label{fig:d6gap}
\end{figure}

To study the effect of the triaxial basal anisotropy, we compute the gaps of the two lower magnon branches as a function of $d_6$, leaving all other parameters fixed. The result is shown in Fig.~\ref{fig:d6gap}, where the gray dashed line corresponds to $d_6=10^{-8}~$meV - the experimental value for $\alpha$-Fe$_2$O$_3$. As we can see, the magnon branch shown in red is independent of $d_6$, which accords with its out-of-plane nature. The branch in blue depends on $d_6$ and can thus be assigned to the in-plane mode. The strong dependence on $d_6$ is due to the exchange enhancement effect discussed above in Eq.~\eqref{eq:model-d6}. However, despite the sensitive dependence of the in-plane magnon gap on $d_6$, only the out-of-plane magnons contribute to AHE as will be shown in Sec.~\ref{sec:non-reciprocal}.

\section{Optical conductivity}\label{sec:non-reciprocal}

We now consider the optical conductivity for the spin Hamiltonian~\Eqref{eq:spin-Hamiltonian}. Let us first discuss the coupling of the system to the electromagnetic field. At low frequencies, an external electric field $\mathbf{E}$ can induce a finite polarization $\mathbf{P}$ in an magnetic insulator~\cite{tokura2014multiferroics}. For our purpose, the polarization is induced microscopically by local charge fluctuations due to virtual hopping of the electrons from absorption of electromagnetic waves~\cite{ng2007power,bolens2018mechanism}. This mechanism differs from charge doping, by which the QAH effect has also been observed in hematite~\cite{galindez2025revealing}. The coupling gives the following term in the Hamiltonian:
\begin{equation}\label{eq:polarization-coupling}
    V = - \mathbf{P}.\mathbf{E},
\end{equation}
where $\mathbf{P}$ can be expressed in terms of spin operators. The form of $\mathbf{P}$ is determined by the system's symmetry~\cite{bolens2018mechanism} and is generally very complex. For hematite we approximate $\mathbf{P}$ along a unit vector $\mathbf{e}_\gamma$ connecting two sites $i,j$ by the following expression:
\begin{equation}\label{eq:polarization}
    \mathbf{P} = \sum_{ij\in \gamma}\alpha_{ij}\left[(\mathbf{S}_i \times \mathbf{S}_j).\mathbf{e}_\gamma\right] \mathbf{e}_\gamma,
\end{equation}
so that Eq.~\eqref{eq:polarization-coupling} can be regarded as a local electric-field-induced DMI. Here $\alpha_{ij}$ are undetermined constants in units of charge times length; we shall estimate its value in Appendix~\ref{sec:estimate}. Note that on a 2D honeycomb lattice with C$_3$ rotation symmetry, Eq.~\eqref{eq:polarization} is exact and can be derived by treating electron hoppings perturbatively using a microscopic Hamiltonian~\cite{bolens2018mechanism}. 

The polarizability is given by the standard Kubo's formula:
\begin{equation}\label{eq:polarizability}
    \chi_{\alpha\beta} (\omega) = \frac{i}{\hbar} \int_0^{\infty} \left\langle [P_\alpha(t), P_\beta(0)]\right\rangle e^{i\omega t} \diff t.
\end{equation}
At low frequencies, the conductivity $\sigma_{\alpha\beta}$ is related to $\chi_{\alpha\beta}$ by~\cite{bolens2018mechanism}:
\begin{equation}\label{eq:conductivity}
     \sigma_{\alpha\beta}(\omega)= -i\omega\chi_{\alpha\beta}(\omega).
\end{equation}

Before proceeding, let us comment on another mechanism for spin-light coupling by electrostriction~\cite{tokura2014multiferroics}, where the external electric field induces a lattice deformation and the associated polarization vector has the form:
\begin{equation}\label{eq:polarization-electrostriction}
    \mathbf{P} = \sum_{i,j \in \gamma} \beta_{ij} (\mathbf{S}_i.\mathbf{S}_j) \mathbf{e}_\gamma.
\end{equation}
However, at $\mathbf{k}=0$ only the $P_z$ component dominates, since low-energy magnons have small momenta and to leading order, one can neglect the unit cell position in \Eqref{eq:polarization-electrostriction}. The summation over C$_3$-symmetric bonds for a given $n$-th N.N. gives either zero or
\begin{equation}
\sum_{\gamma \in \text{n-th N.N.}} \mathbf{e}_\gamma \propto \mathbf{e}_z.
\end{equation}
Such a $\mathbf{P}$ would only give a non-zero $\sigma_{zz}$, and therefore the QAH effect cannot arise from this mechanism. We note that generally $\mathbf{e}_\gamma$ does not need to be along the bonds. But our conclusion still holds since $\mathbf{e}_\gamma$ still comes in C$_3$-symmetric triplets.

In what follows, we shall include in Eq.~\eqref{eq:polarization} up to the third N.N. bond. The N.N. bonds between A and B, and C and D are along $\mathbf{e}_z$, which only contributes to $\chi_{zz}$ as can be seen from \eqref{eq:polarizability}. The 2nd N.N. bonds are between sublattices with the same magnetic moment (A-D and B-C) which produces a real constant polarizability and an imaginary conductivity from Eq.~\eqref{eq:conductivity}. Thus only the third N.N. bonds (A-B and C-D) contribute to the real part of conductivity. The corresponding bond vectors $\mathbf{e}_\gamma$ can be given in terms of Fe atom positions inside a unit cell $\mathbf{r}_n$ where $n=A,...,D$ is the sublattice index, and the primitive lattice vectors $\mathbf{a}_i$. For example from A to B:
\begin{equation}
    \mathbf{e}^{(3)}_{i} \propto \mathbf{r}_B -\mathbf{r}_A - \mathbf{a}_i, \ i=1,2,3.
\end{equation}
The bond vectors from C to D are the same. The lattice vector values are presented in Ref.~\cite{hoyer2025altermagnetic}. The values of $\mathbf{r}_{A,B}$ are given in terms of fractional coordinates:
\begin{align}
    &\mathbf{r}_A = \left(\frac{1}{6}-\delta_{\text{Fe}},\frac{1}{6}-\delta_{\text{Fe}},\frac{1}{6}-\delta_{\text{Fe}}\right), \\
    &\mathbf{r}_B = \left(\frac{1}{3}+\delta_{\text{Fe}},\frac{1}{3}+\delta_{\text{Fe}},\frac{1}{3}+\delta_{\text{Fe}}\right),
\end{align}
where $\delta_{\text{Fe}}\approx 0.02165$. The vector $\mathbf{a}_1$ is given in Cartesian coordinates by
\begin{align}
    \mathbf{a}_1 = \left(-\frac{a_c}{2\sqrt{3}},\frac{a_c}{2}, \frac{c_c}{3}\right),
\end{align}
and $\mathbf{a}_{2,3}$ are obtained from $\mathbf{a}_1$ by successive C$_3$ rotations with $a_c = 5.0342~\r{A} $ and $c_c = 13.7519~\r{A}$ being conventional lattice vector lengths. 

We now discuss the conductivity. Let us first consider general properties of the conductivity tensor. As follows from Onsager's relation for the conductivity~\cite{rao2024tunable} (see also Sec.~101 in Ref.~\cite{landau1984electrodynamics}), the conductivity tensor can be decomposed into symmetric and anti-symmetric components:
\begin{equation}
    \sigma_{\alpha\beta}(\omega;\mathbf{m}) = \sigma_{\alpha\beta}^{\text{(s)}}(\omega;\mathbf{m}) + \sigma_{\alpha\beta}^{\text{(as)}}(\omega;\mathbf{m}),
\end{equation}
where $\omega$ is frequency and $\mathbf{m}$ is the canted total moment. The non-reciprocal conditions are given by:
\begin{equation}\label{eq:non-reciprocal}
    \sigma_{\alpha\beta}^{\text{(s)}}(\omega;-\mathbf{m}) = \sigma_{\alpha\beta}^{\text{(s)}}(\omega;\mathbf{m}); \ \sigma_{\alpha\beta}^{\text{(as)}}(\omega;-\mathbf{m}) = -\sigma_{\alpha\beta}^{\text{(as)}}(\omega;\mathbf{m}).
\end{equation}
In other words, $\mathbf{m}$ is a measure of the non-reciprocity due to DMI.

In Appendix~\ref{sec:magnon}, we evaluate equations \eqref{eq:polarizability} and \eqref{eq:conductivity} using the Matsubara Green's functions method, taking into account only single-magnon contributions. As shown in Appendix~\ref{sec:magnon}, the polarizability tensor contains simple poles at the magnon gaps $\chi_{\alpha\beta}(\omega)\sim (\omega - \Delta + i \delta)^{-1}$, where $\delta$ is a positive infinitesimal at zero temperature. At non-zero temperatures, $\delta$ acquires a finite value due to effects of thermal broadening. Physically $\delta$ is determined by the magnon-phonon and magnon-magnon interactions. A microscopic calculation of $\delta$ is outside the scope of the present paper. However, to show the non-zero QAH effect, we set $\delta = 0.01~$meV in numerical simulations. We note that at finite temperatures, magnon-phonon and magnon-magnon scattering very likely give rise to a temperature-dependent linewidth, which broadens with temperature.

Before presenting the results, we comment on the choice of \eqref{eq:polarization} for the polarization operator. We expect the general conclusions to not depend on the specific form of $\mathbf{P}$: Eq.~\eqref{eq:polarization} contains a general superposition of $S^x,S^y,S^z$ as $\mathbf{e}_\gamma$ does not represent any high-symmetry axes. Indeed we check that changing $\mathbf{e}_\gamma$ does not affect qualitatively our conclusions. This suggests that the qualitative properties of conductivities are determined by system symmetry, reflected in the summation of C$_3$-symmetric bonds in \Eqref{eq:polarization}.

\subsection{Longitudinal conductivity}

\begin{figure}
    \centering
    \includegraphics[width=1\linewidth]{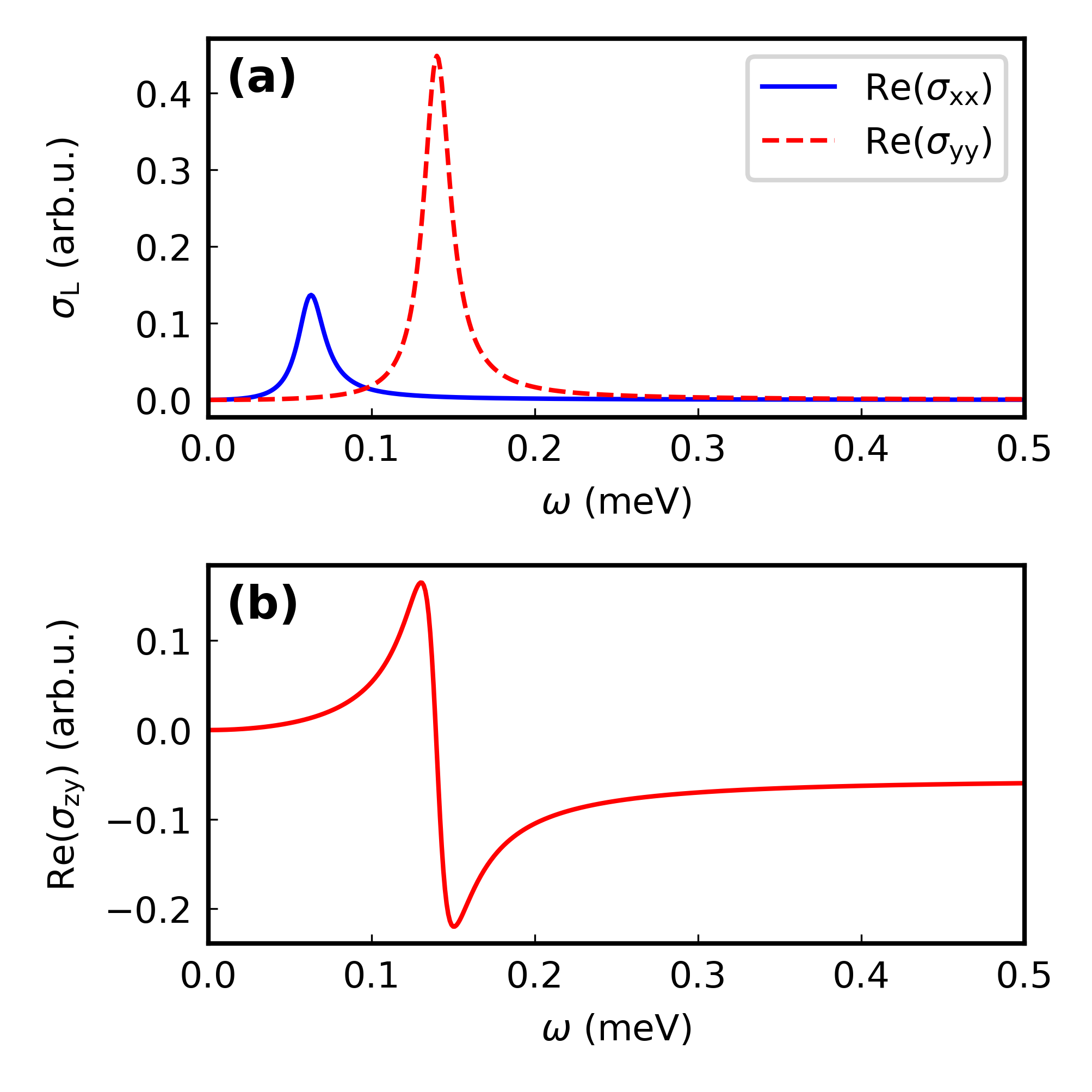}
    \caption{Real part of conductivities of the spin Hamiltonian~\Eqref{eq:spin-Hamiltonian} as a function of frequency $\omega$. (a) Longitudinal conductivities where $\Re\sigma_{xx}$ and $\Re\sigma_{yy}$ couple to the lower and higher magnon branches in Fig.~\eqref{fig:LSW}(b) respectively. (b) The non-zero antisymmetric conductivity component satisfying \Eqref{eq:AHE}, coupling to the higher magnon branch. The other off-diagonal conductivities are zero.}
    \label{fig:conductivity}
\end{figure}

For numerical calculations, we take $\alpha_{ij}=1$.  It is demonstrated in Appendix~\ref{sec:estimate} that $\alpha_{ij}^2/A\hbar  = C(e^2/\hbar)$ and the conductivity has the correct dimension $e^2/\hbar$.  We shall take $C=0.1$ for numerical simulations which is large for hematite according to our estimates in Appendix~\ref{sec:estimate}. In general we note that $C$ can only be determined from microscopics.  
We first compute the symmetric components $\Re\sigma_{\alpha\beta}^{(\text{s})}$ as a function of frequency. We find that the principle axes of the conductivities are determined by the direction of $\mathbf{m}$ and the $c$-axis. Since $\mathbf{m}\parallel \mathbf{e}_x$ in our case, this makes the $xyz$-axes the principle axes and the symmetric conductivities are diagonal. As shown in Fig.~\ref{fig:conductivity}(a), $\Re \sigma_{xx}^{(\text{s})}$ has a peak corresponding to the in-plane pseudo-Goldstone gap, and $\Re \sigma_{yy}^{(\text{s})}$ couples to the second magnon branch due to out-of-plane oscillations. $\Re \sigma_{zz}^{(\text{s})}$ has the same peak location as $\Re \sigma_{yy}^{(\text{s})}$ albeit with a different intensity and is not shown here.

\subsection{Hall conductivity}

A finite DMI induces canting of the magnetic moments and a total net moment $\mathbf{m}$. For our given $\mathbf{m}\parallel \mathbf{e}_x$, we compute the anomalous Hall conductivity, which to leading order in $\mathbf{m}$ has the following form:
\begin{equation}\label{eq:AHE}
    \sigma_{\alpha\beta}^{\text{as}}(\omega) \propto \varepsilon_{\alpha\beta \gamma} m^\gamma,
\end{equation}
and satisfies the non-reciprocal condition Eq.~\ref{eq:non-reciprocal}. We confirm that in our system with $\mathbf{m} \parallel \mathbf{e}_x$, only $\sigma_{zy} = - \sigma_{yz}$ are non-zero for off-diagonal conductivity components. The result is plotted in Fig.~\ref{fig:conductivity}(b).

Only the out-of-plane magnon mode contributes to the Hall conductivity. Its peak intensity also depends on the DMI strength. In particular, we verify that $\sigma^{(\text{as})}$ vanishes in the absence of DMI. This is expected because the out-of-plane mode is governed by the DMI interactions, which is responsible for TRS breaking by inducing a small $\mathbf{m}$. 

Note that in Fig.~\ref{fig:conductivity}(b) $\lim_{\omega \rightarrow \infty}\sigma_{zy}(\omega) = \text{const.}$, which is an artifact of the single-magnon approximation as can be seen as follows. The polarizability can be expressed as a linear superposition of magnon Green's functions so at $\omega \rightarrow \infty$, $ \chi (\omega)\sim 1/\omega$. Since the expression for the conductivity~\Eqref{eq:conductivity} contains an additional multiplication by $\omega$, $\sigma_{zy}$ tends to a constant value at large $\omega$ as a result. However, the unphysical constant value is an artifact of the single-magnon approximation, as at higher frequencies multiple magnon contributions become important. Furthermore Eq.~\eqref{eq:conductivity} relating $\chi$ and $\sigma$ holds only at small frequencies.

\section{Non-reciprocal circulator}\label{sec:circulators}

\begin{figure}
    \centering
    \includegraphics[width=0.9\linewidth]{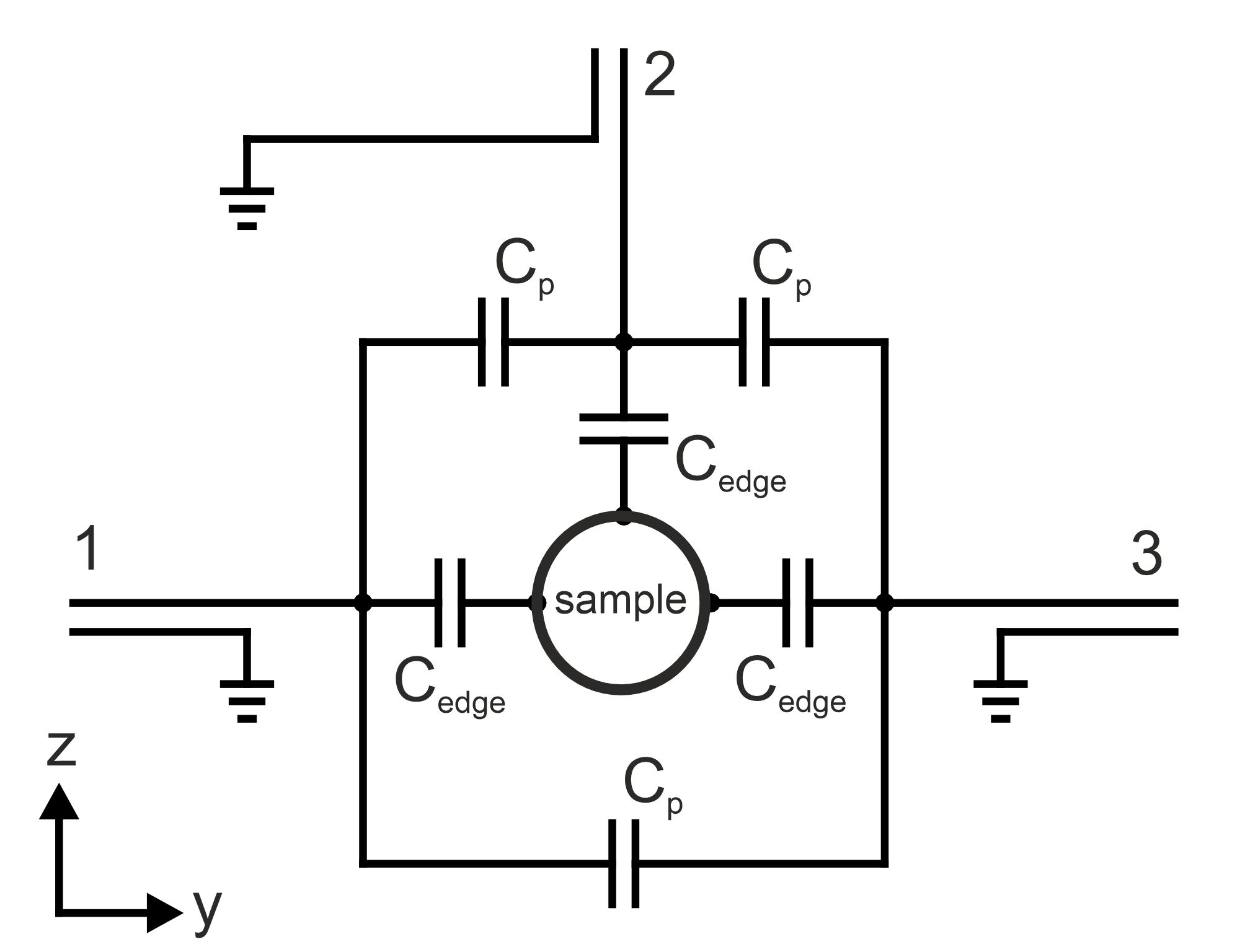}
    \caption{Scheme of a three-port circulator with the sample material in the center. The circulator ports 1, 2, and 3 are capacitively coupled via $C_\mathrm{edge}$ to the sample (hematite).  The term $C_\mathrm{p}$ describes the parasitic coupling between the port arms. 
    }
    \label{fig:circuitDrawing}
\end{figure}

\begin{figure}
    \centering
    \includegraphics[width=1\linewidth]{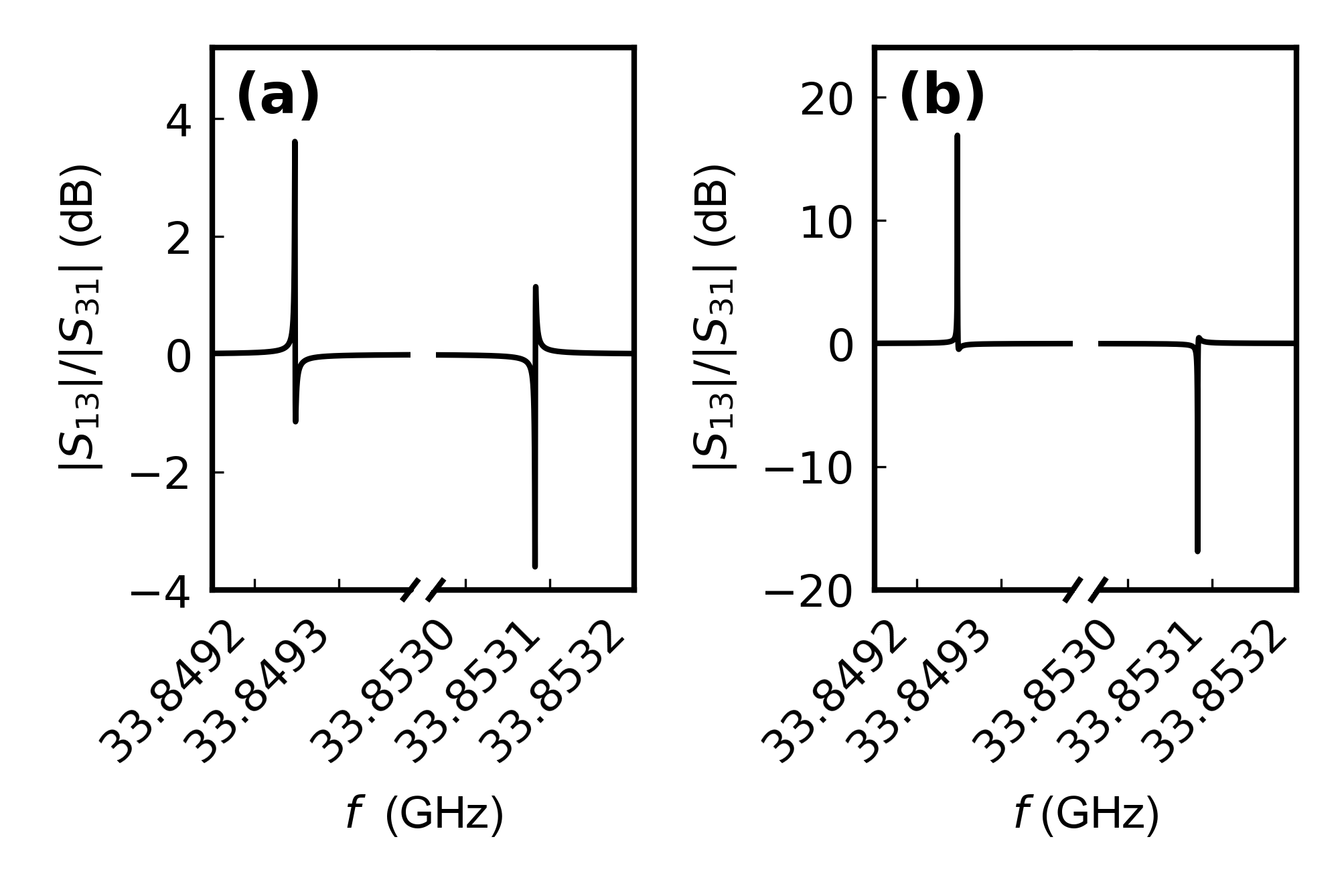}
    \caption{Simulated nonreciprocal response $|S_\mathrm{13}|/|S_\mathrm{31}|$  versus frequency for a hematite-based circulator with $C_\mathrm{edge}~=~67.35~\mathrm{fF}$ and $C_\mathrm{p}~=~15~\mathrm{fF}$, showing sharp periodic resonances. (a) Non-dissipative case with resonance amplitude $\sim4~\mathrm{dB}$. (b) Dissipative case ($R_\mathrm{dissipative} = 50~\Omega$) exhibiting enhanced resonance amplitude $\sim17~\mathrm{dB}$ and narrower linewidth. Plotes of individual scattering-parameters $|S_{13}|$ and $|S_{31}|$ are shown in Appendix~\ref{sec:circulator}.}
    \label{fig:circularity}
\end{figure}

We model the circularity of a potential hematite-circulator using the scattering matrix formalism of A.~Mahoney \textit{et al.}\,\cite{mahoney2017chip} (Appendix~\ref{sec:circulator}), which has four free parameters $\sigma_\mathrm{yz}$, $C_\mathrm{edge}$, $C_\mathrm{p}$, and $R_\mathrm{dissipative}$ (see Fig.~\ref{fig:circuitDrawing}). 
The behavior of an ideal circulator can be described by the scattering-matrix
\begin{equation}\label{eq:scatteringMatrixCirculator}
    \mathbf{S} = \begin{pmatrix} 0 & 1 & 0 \\0 & 0 & 1 \\1 & 0 & 0  \end{pmatrix}, 
\end{equation}
meaning that a signal entering the circulator at port 1 exits at port 3, a signal entering port 2 exits at port 1, and a signal entering port 3 exits at port 2 \cite{viola2014}. The circularity of a real circulator can be described by dividing the absolute elements of the scattering-matrix, e.g. $|S_\mathrm{13}|/|S_\mathrm{31}|$. A detailed discussion of the underlying circulator methodologies can be found in Ref.~\cite{viola2014}.
In our model (see Fig.~\ref{fig:circuitDrawing}), $\sigma_\mathrm{yz}$ and $R_\mathrm{dissipative}$ are materials parameters, while $C_\mathrm{edge}$ and $C_\mathrm{p}$ are set by the circuit design. Specifically, the conductivity $\sigma_\mathrm{yz}$ links to the sections above and therefore to the materials properties of $\alpha$-Fe$_2$O$_3$, $C_\mathrm{edge}$ is the capacitive coupling to the hematite, $C_\mathrm{p}$ is the parasitic coupling between the contact arms, and $R_\mathrm{dissipative}$ is a phenomenological parameter that describes the dissipation caused by imperfections in the system \cite{mahoney2017chip}.

When we choose $C_\mathrm{p}$ and $C_\mathrm{edge}$ such that the circularity is maximized, we see repeating comb-like resonances over a wide range of frequencies reaching circularities of up to $17~\mathrm{dB}$ (Fig.~\ref{fig:circularity}).  The particular resonance frequencies are primarily set by the edge-channel phase in the circulator, i.e. by the condition that the dimensionless phase  $\varphi = \omega C_{\mathrm{edge}}/\sigma_{\mathrm{yz}}$ (Appendix~\ref{sec:circulator}) hits values that make the 3‑port admittance (and therefore the scattering matrix) resonant. In the circuit model, $R_\mathrm{dissipative}$ enters the total admittance as a phenomenological loss parameter and it modifies both resonance contrast and linewidth. For the data as in Fig.~6, $R_\mathrm{dissipative}$ increases the peak circularity but narrows the resonances [cf. Figs.~6(a) and 6(b)]. Generally, the computed values of the ratio $|S_\mathrm{13}|/|S_\mathrm{31}|$ compare reasonably well with work by A.~Mahoney \textit{et al.}\,, e.g. on circulator geometries based on topological insulators \cite{mahoney2017chip, mahoney2017zero}. However, future work will need to clarify how the capacitances of a specific circuit with the impact of a substrate and fabrication uncertainties etc. can be optimized to result in larger widths of the resonances.

\section{Conclusions}
Our calculations and modeling show that an easy-plane antiferromagnetic insulator, such as hematite ($\alpha$-Fe$_2$O$_3$), with finite canting of the magnetic sublattices induced via DMI provides a perspective for non-reciprocal circulators due to a finite anomalous Hall conductivity. The operation frequency can be tuned via the DMI and magnetic anisotropy in this system, such that by careful material engineering, e.g. via doping and epitaxial strain tuning of the thin films properties, the operational frequency range of the circulator can be adjusted. Also the temperature range of the weak ferromagnetic phase can be tuned in hematite via doping and thin film growth conditions~\cite{besser1967dopedhematite,Park2013StrainMorin,Mibu2017Thickness,jani2021hydrogenhematite,scheufele2023hematite}. While we here focus on $\alpha$-Fe$_2$O$_3$ as a prototype example, it is important to highlight that our model and description are universal, such that other antiferromagnetic insulators can also be considered. For example, for operation at room temperature the orthoferrites (RFeO$_3$, R representing a rare earth element)~\cite{nordlander2022,li2025}, exhibiting magnetic ordering temperatures above room temperature and strong DMI, can be considered an exciting alternative. For low temperature operation, the hexagonal manganites (RMnO$_3$)~\cite{lilienblum2015,nordlander2022} can be a viable material class. As far as the simulation of the non-reciprocal circulator are concerned,  we find narrow frequency bands for circulator operation. The narrow resonances may be useful for frequency-selective applications if the operating frequency can be tuned and stabilized within the resonance window. We point out that their applicability will depend on whether that window is robust to temperature-dependent magnon damping and to fabrication-induced parameter variations. However, the operation bandwidth is mainly determined by $R_\mathrm{dissipative}$ and might be adjusted  for a more broadband application.

In summary, we here provide a detailed discussion of the non-reciprocal response of the optical conductivity in hematite in the canted easy-plane antiferromagnetic phase and discuss a circulator design in the GHz frequencies for future applications. This approach can be extended to other non-collinear antiferromagnetic insulators and opens up a new avenue to explore for non-reciprocal devices in the microwave domain.


\acknowledgments{P.R. thanks Rhea Hoyer for a discussion on the hematite crystal structure. This research is part of the Munich
Quantum Valley (K1), which is
supported by the Bavarian state government with funds
from the Hightech Agenda Bayern Plus.}

\section*{Data Availability}
The Mathematica notebook for numerically calculating the linear spin wave spectrum and conductivity is available at \cite{code}.

\appendix

\section{Linear spin wave theory}\label{sec:LSW}

Let us briefly review the linear spin wave theory which we implement numerically. To obtain the magnon spectrum, the spin operators are rotated into the local axis along the directions of sublattice magnetic moment which we shall denote by $z'$ (note for different sublattices $z'$-axes are different). The HP approximation then expresses the spin Hamiltonian in terms of magnon operators $a_i, \ i=A,B,C,D$ as follows:
\begin{equation}
 S^{+}_i = S^{x'}_i +iS^{y'}_i= \sqrt{2S-a_i^\dagger a_i}a_i, \  S^{z'}_i = S-a_i^\dagger a_i.
\end{equation}
Subsequently we expand the results in a series of $1/S$. We verify that terms linear in the magnon operators $a_i$ cancel for a given classical GS. The quadratic terms give the magnon Hamiltonian which can be written as a BdG Hamiltonian in the particle-hole basis $\Psi_\mathbf{k} = (a_{A,\mathbf{k}},...,a_{D,\mathbf{k}},a_{A,-\mathbf{k}}^\dagger,...,a_{D,-\mathbf{k}}^\dagger)$:
\begin{equation}\label{eq:magnon-Hamiltonian}
\begin{split}
    H_{2} =&\frac{1}{2}\sum_\mathbf{k} \Psi_\mathbf{k}^\dagger H_{\text{BdG}}(\mathbf{k})\Psi_\mathbf{k} \\
 =& \frac{1}{2}\sum_{i,j} \left[\varepsilon_{ij}(\mathbf{k}) a_{i,\mathbf{k}}^\dagger a_{j,\mathbf{k}} + \Delta_{ij}(\mathbf{k}) a_{i,\mathbf{k}}^\dagger a_{j,-\mathbf{k}}^\dagger +\text{h.c.} \right].
\end{split}
\end{equation}
$H_{\text{BdG}}(\mathbf{k})$ has the following block structure in particle-hole space:
\begin{equation}
     H_{\text{BdG}}(\mathbf{k}) =\begin{bmatrix} \varepsilon(\mathbf{k}) & \Delta (\mathbf{k}) \\ \Delta^\dagger (\mathbf{k}) & \varepsilon^T(-\mathbf{k})
     \end{bmatrix}.
\end{equation}
From the boson commutation relations and hermiticity of $H_{\text{BdG}}(\mathbf{k})$ we get:
\begin{equation}\label{eq:Hamiltonian-identity}
\Delta^T(\mathbf{k})= \Delta(-\mathbf{k}), \  \varepsilon^\dagger(\mathbf{k}) = \varepsilon(\mathbf{k}).
\end{equation}

The BdG Hamiltonian is diagonalized by performing the following Bogoliubov transformation into new quasi-particle operators $b_{n,\mathbf{k}}, \ n=1,...,4$:
\begin{equation}\label{eq:bogoliubov}
    \Psi_\mathbf{k} = U_\mathbf{k} \widetilde{\Psi}_\mathbf{k}, \ \widetilde{\Psi}_\mathbf{k} = \left(b_{1,\mathbf{k}}, ...,b_{4,\mathbf{k}}, b_{1,-\mathbf{k}}^\dagger, ..., b_{4,-\mathbf{k}}^\dagger\right).
\end{equation}
The standard boson commutation relation can be written as:
\begin{equation}
    [\Psi_{\mathbf{k},i}, \Psi^\dagger_{\mathbf{k},j}] = (\tau^z)_{ij}, \ [\widetilde{\Psi}_{\mathbf{k},m}, \widetilde{\Psi}_{\mathbf{k},n}^\dagger] =(\tau^z)_{mn},
\end{equation}
where $\tau^z$ is the third Pauli matrix in particle-hole space. Substituting \eqref{eq:bogoliubov} we obtain the paraunitary condition on $U_\mathbf{k}$:
\begin{equation}\label{eq:paraunitary}
    U_\mathbf{k}^{-1} =  \tau^z U_\mathbf{k}^\dagger \tau^z.
\end{equation}
The Hamiltonian \eqref{eq:magnon-Hamiltonian} becomes:
\begin{equation}\label{eq:magnon-Hamiltonian-1}
    H_2 = \frac{1}{2}\sum_\mathbf{k} \widetilde{\Psi}_\mathbf{k}^\dagger \tau^z U_\mathbf{k}^{-1} \tau^z H_{\text{BdG}}(\mathbf{k})U_\mathbf{k}\widetilde{\Psi}_\mathbf{k}.
\end{equation}
Therefore, $U_\mathbf{k}$ diagonalizes $\tau^z H_{\text{BdG}}(\mathbf{k})$. 

Equation~\eqref{eq:bogoliubov} and $U_\mathbf{k}$ can be written explicitly:
\begin{align}\label{eq:bogoliubov-1}
&a_{i,\mathbf{k}} = \sum_n\left[u_{in}(\mathbf{k})b_{n,\mathbf{k}} + v_{in}^*(-\mathbf{k})b_{n,-\mathbf{k}}^\dagger\right]; \\
& U_\mathbf{k} = \begin{bmatrix} u(\mathbf{k}) & v^*(-\mathbf{k}) \\ v(\mathbf{k}) & u^*(-\mathbf{k}) \end{bmatrix}.
\end{align}
The paraunitary condition \eqref{eq:paraunitary} imposes the normalization condition:
\begin{equation}
    \sum_i \left[ u_{in}^* (\mathbf{k})u_{im}(\mathbf{k}) - v_{in} (\mathbf{k})v_{im}^* (\mathbf{k})\right]=\delta_{mn}.
\end{equation}

We now show that $\tau^z H_{\text{BdG}}(\mathbf{k})$ can be diagonalized into the following particle-hole symmetric form:
\begin{equation}\label{eq:magnon-Hamiltonian-diagonal}
\begin{split}
  &U_\mathbf{k}^{-1} \tau^z H_{\text{BdG}}(\mathbf{k})U_\mathbf{k} \\
  =&  \diag\{E_{1}(\mathbf{k}), ..., E_{4}(\mathbf{k}), -E_{1}(-\mathbf{k}),..., -E_{4}(-\mathbf{k})\}.
\end{split}
\end{equation}
First note that the eigenvectors of $\tau^z H_{\text{BdG}}(\mathbf{k})$ are column vectors of $U_\mathbf{k}$ in Equation~\eqref{eq:bogoliubov-1}. We shall assume that the first four column vectors $[u_{n}(\mathbf{k}),v_{n}(\mathbf{k})]$ have eigenvalues $E_n(\mathbf{k})$:
\begin{equation}
    \begin{bmatrix} \varepsilon(\mathbf{k}) & \Delta (\mathbf{k}) \\ -\Delta^\dagger (\mathbf{k}) & -\varepsilon^T(-\mathbf{k}) \end{bmatrix}\begin{bmatrix}u_{n}(\mathbf{k}) \\ v_n(\mathbf{k}) \end{bmatrix} = E_n(\mathbf{k}) \begin{bmatrix}u_{n}(\mathbf{k}) \\ v_n(\mathbf{k}) \end{bmatrix}.
\end{equation}
Taking the complex conjugate and $\mathbf{k}\rightarrow -\mathbf{k}$ then making use of the identity \eqref{eq:Hamiltonian-identity}, it can be shown that:
\begin{equation}
    \begin{bmatrix} \varepsilon(\mathbf{k}) & \Delta (\mathbf{k}) \\ -\Delta^\dagger (\mathbf{k}) & -\varepsilon^T(-\mathbf{k}) \end{bmatrix}\begin{bmatrix}v_{n}^*(-\mathbf{k}) \\ u_n^*(-\mathbf{k}) \end{bmatrix} =- E_n(-\mathbf{k}) \begin{bmatrix}v_{n}^*(-\mathbf{k}) \\ u_n^*(-\mathbf{k}) \end{bmatrix}.
\end{equation}
Thus the last four column vectors of $U_\mathbf{k}$ are indeed eigenvectors with eigenvalues $-E_n(-\mathbf{k})$. This proves our statement. 

Finally, substituting Eqs.~\eqref{eq:bogoliubov} and \eqref{eq:magnon-Hamiltonian-diagonal} into \eqref{eq:magnon-Hamiltonian-1}, we obtain:
\begin{equation}
    H_2 = \sum_{n,\mathbf{k}} E_n(\mathbf{k}) \left( b_{n,\mathbf{k}}^\dagger b_{n,\mathbf{k}} + \frac{1}{2} \right).
\end{equation}
Thus $E_n(\mathbf{k})$ are the magnon dispersions.

\section{Estimates of the constant $\alpha_{ij}$}\label{sec:estimate}
As mentioned in Sec.~\ref{sec:non-reciprocal}, the proportionality constant $\alpha_{ij}$ in the polarization~\eqref{eq:polarization} has the dimension charge times length, i.e. C$\cdot$m. According to the Kubo's formula this gives for the unit of conductivity:
\begin{equation}
    \sigma \sim \frac{\omega \alpha_{ij}^2}{A\hbar } \int \langle P^2\rangle e^{i\omega t}\diff t \sim \frac{\alpha_{ij}^2}{A\hbar}, 
\end{equation}
where $A$ is the unit cell area of the two-dimensional sample. Thus $\sigma$ is given in units of $e^2/\hbar$ as it should. Here the dimension of $\omega$ does not enter. This is clear from the fact that the single-magnon Green's functions give $\omega \langle P^2\rangle \sim \omega (\omega-E)^{-1}$.

Estimating $\alpha_{ij}$ directly is difficult, since hematite is not ferroelectric. Instead we shall use the magneto-electric effect which states for the polarization density $\mathbf{p}$:
\begin{equation}
    \mathbf{p} = \gamma \mathbf{B}.
\end{equation}
Let us consider hematite in the AFM phase with the magnetic moments along the $c$-axis, and apply an in-plane magnetic field. The field introduces an canting that can be estimated by considering a single sublattice $A$ and the dominant 3rd N.N. coupling to $B$. Assuming the magnetic moments to be almost parallel to $c$ but canted away towards $\mathbf{B}$ by angle $\theta$, the classical energy per sublattice is:
\begin{equation}
    E =- 3J_3S^2 \cos 2 \theta  - 2\mu_B S \sin \theta.
\end{equation}
Here the factor of three is due to $3$ third N.N. pairs and $g=2$. Minimizing $E$ gives:
\begin{equation}
    \sin\theta \approx \theta =\frac{2\mu_B B }{3J_3 S}.
\end{equation}
Now averaging over the polarization density directly using \eqref{eq:polarization} gives:
\begin{equation}
    |\mathbf{p}| \sim \frac{ \alpha S^2\theta }{V},
\end{equation}
where $V\approx 3\times 10^{-28}~$m is the volume of the 3D unit cell. This gives for the magneto-electric (ME) coefficient:
\begin{equation}
   \alpha =  \frac{3J_3 \gamma V }{2\mu_B S}.
\end{equation}
The value $\alpha^2/\hbar A \approx 0.08 (e^2/\hbar)$ corresponds to $a \mu_0 = 600~$ps/m, where $A = a c$ the 2D unit cell projected to the $yz$-plane. Although this is far larger than the typical value range $a\mu_0 \sim 1-10~$ps/m (for example for Cr$_2$O$_3$, $a\mu_0 \approx 4.3$ps/m at $260~$K~\cite{schoenherr2017magnetoelectric}) and only slightly below the largest ME coefficient $730~$ps/m for single-phase crystals~\cite{rivera2009short}, we note that our estimation method is very crude and the ME coefficient is highly sensitive on experimental conditions such as the temperatures. A much larger ME coefficient has been achieved for heterostructures, e.g. see Refs.~\cite{eerenstein2007giant,pertsev2009strong}.

\section{Computation of polarizability}\label{sec:magnon}

To compute the polarizability \eqref{eq:polarizability} it is convenient to use the Matsubara Green's functions. In imaginary time $\tau$, the polarizability is given by the following:
\begin{equation}\label{eq:polarizability-Matsubara}
    \chi_{\alpha\beta}(\omega) =\lim_{i\omega \rightarrow \omega + i\delta} \int \left\langle \text{T}_\tau P_\alpha(\tau)P_\beta(0)\right\rangle  e^{i\omega \tau} \diff \tau,
\end{equation}
where $\delta$ is a positive infinitesimal from analytical continuation. 

We expand the polarization vector \eqref{eq:polarization} in terms of magnon operators $\Psi_i$ for a given classical ground state. In what follows we shall only keep the terms linear in $\Psi_i$:
\begin{equation}\label{eq:polarization-magnon}
    P_{\alpha}(\tau) =\text{const}. + \sum_i c_{\alpha i} \Psi_i (\tau).
\end{equation}
As we shall see below, since the polarizability $\chi_{\alpha\beta}$~\eqref{eq:polarizability-Matsubara} is evaluated at $\mathbf{k}=0$, the position dependence of $\Psi_i$ can be neglected. In substituting \eqref{eq:polarization-magnon} into \eqref{eq:polarization}, we obtain:
\begin{equation}
     \chi_{\alpha\beta}(\omega) =\lim_{i\omega \rightarrow \omega + i\delta} \sum_{i,j}c_{\alpha i} c_{\beta j}^*\int e^{i\omega \tau} \diff \tau\left\langle \text{T}_\tau \Psi_i(\tau)\Psi_j^\dagger (0)\right\rangle .
\end{equation}
Here we neglected the constant part in \eqref{eq:polarization-magnon}, which to second order gives averages of the form $\left\langle \text{T}_\tau \Psi_i(\tau)\Psi_j^\dagger (\tau)\right\rangle, \left\langle \text{T}_\tau \Psi_i(0)\Psi_j^\dagger (0)\right\rangle$ and does not contribute to the finite frequency response. Thus the truncation of \eqref{eq:polarization-magnon} to linear order means only single-magnon states contribute. As mentioned earlier, in evaluating at $\mathbf{k}=0$, the unit cell position of $\Psi_i$ does not enter. 

The magnon Green's functions $\left\langle \text{T}_\tau \Psi_i(\tau)\Psi_j^\dagger (0)\right\rangle$ can be obtained from the quadratic magnon Hamiltonian as follows:
\begin{align}
    &-\left\langle \text{T}_\tau \Psi_i(\tau)\Psi_j^\dagger (0)\right\rangle=\left( -\frac{\p}{\p\tau} \tau^z- H_{\text{BdG}} \right)_{ij}^{-1} \\
    =& -\begin{bmatrix}
        \left\langle \text{T}_\tau b(\tau)b^\dagger (0)\right\rangle  &\left\langle \text{T}_\tau b(\tau)b (0)\right\rangle \\ 
        \left\langle \text{T}_\tau b^\dagger(\tau)b^\dagger (0) \right\rangle & \left\langle \text{T}_\tau b^\dagger(\tau)b (0)\right\rangle
    \end{bmatrix}_{ij}.
\end{align}
Transforming to frequency space and performing the analytical continuation $i\omega \rightarrow \omega + i\delta$ gives the polarizability~\eqref{eq:polarizability} and therefore the conductivity \eqref{eq:conductivity} in the main text. For numerical calculations, we set $\delta =0.01~$meV. 

\section{Circulator scattering parameters}\label{sec:circulator}

Following the model of Mahoney~et~al.\cite{mahoney2017chip}, the scattering matrix of a three-port circulator is calculated from the total admittance matrix $\mathbf{Y}_\mathrm{total}$ and the feedline impedance $Z_0 = 50~\Omega$:

\begin{equation}\label{eq:scatteringMatrix}
    \mathbf{S}(\omega) = \left( \frac{1}{Z_0}\mathbb{I}_3+\mathbf{Y}_\mathrm{total} \right)^{-1} \left( \frac{1}{Z_0}\mathbb{I}_3-\mathbf{Y}_\mathrm{total} \right),
\end{equation}

where $\mathbb{I}_3$ is the $3\times3$ identity matrix. The total admittance combines the edge channel admittance with the dissipative resistance $R_\mathrm{dissipative}$ and the parasitic capacitive coupling $C_\mathrm{p}$ between the contact arms:

\begin{equation}\label{eq:totalAdmittance}
    \mathbf{Y}_{\mathrm{total}} = (\mathbb{I}_3 + R_\mathrm{dissipative} \mathbf{Y}_{\mathrm{edge}})^{-1} \mathbf{Y}_{\mathrm{edge}} 
+ i \omega C_\mathrm{p} \begin{pmatrix} 
2 & -1  & -1  \\ 
-1  & 2  & -1  \\ 
-1  & -1  & 2  
\end{pmatrix}.
\end{equation}

The edge channel admittance matrix $\mathbf{Y}_{\mathrm{edge}}$ has a structure reflecting the threefold rotational symmetry:

\begin{equation}\label{eq:edgeAdmittance}
    \mathbf{Y}_{\mathrm{edge}}(\omega) = \begin{pmatrix} 
ia & b & -b^* \\ 
-b^* & ia & b \\ 
b & -b^* & ia 
\end{pmatrix},
\end{equation}

with diagonal element $a$ and off-diagonal elements $b$ given by:

\begin{equation}\label{eq:diagonalElement}
    a = \frac{2\sigma_{\mathrm{yz}} \sin\varphi}{1 + 2\cos\varphi},
\end{equation}

\begin{equation}\label{eq:offDiagonalElement}
    b = \sigma_{\mathrm{yz}} \frac{-1 + e^{-i\varphi}}{1 + 2\cos\varphi},
\end{equation}

where $\varphi = \omega C_{\mathrm{edge}}/\sigma_{\mathrm{yz}}$ is the dimensionless phase parameter characterizing the edge channel response.

\begin{figure}
    \centering
    \includegraphics[width=1\linewidth]{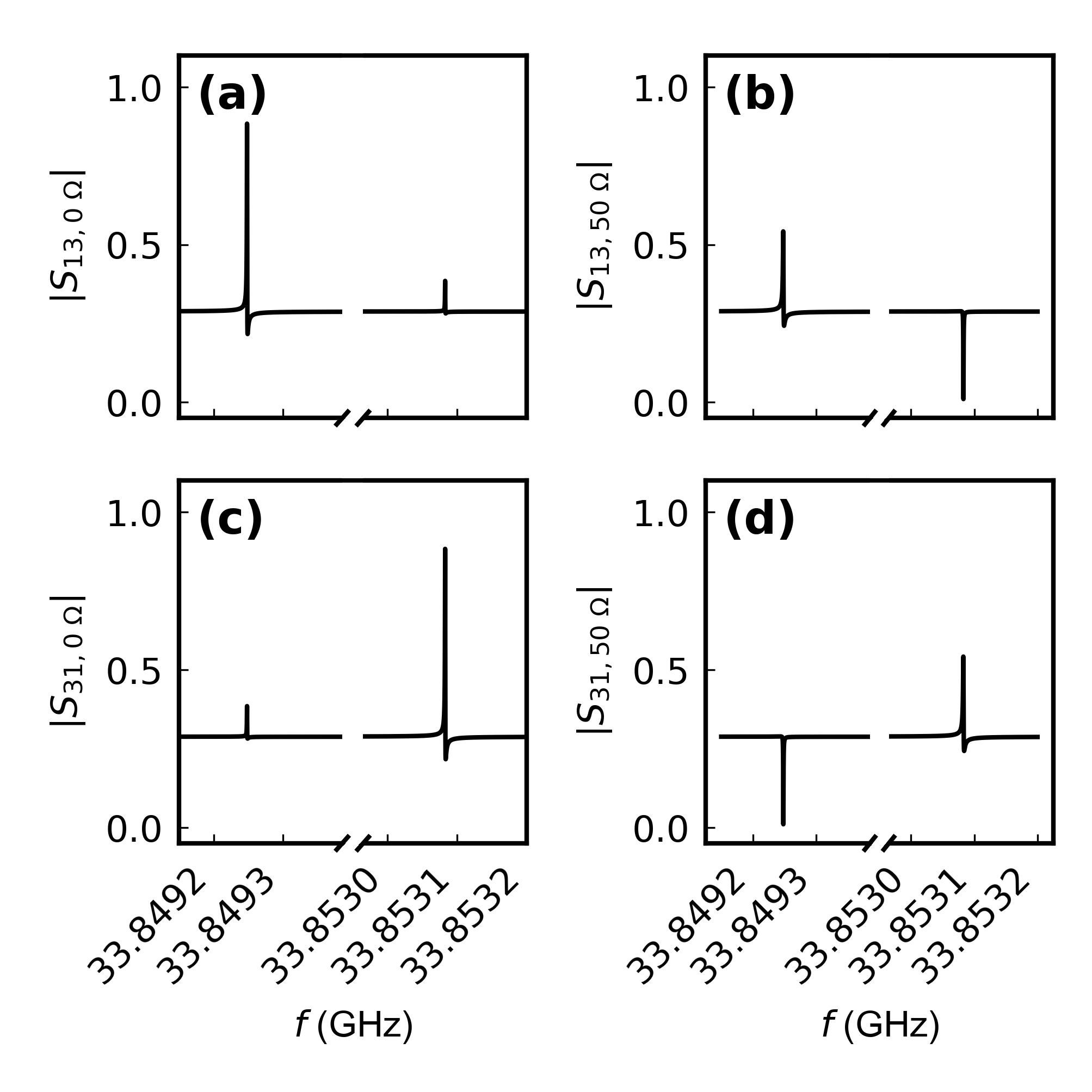}
    \caption{Individual scattering matrix elements $|S_{13}|$ and $|S_{31}|$ for a hematite circulator with $C_\mathrm{edge}~=~67.35~\mathrm{fF}$ and $C_\mathrm{p}~=~15~\mathrm{fF}$. (a,c) Non-dissipative case. (b,d) Dissipative case with $R_\mathrm{dissipative}~=~50~\Omega$.}
    \label{fig:circularityAppendix}
\end{figure}

Figure~\ref{fig:circularityAppendix} shows the scattering parameters $|S_{13}|$ and $|S_{31}|$ for a hematite-based circulator, demonstrating sharp resonances characteristic of non-reciprocal transmission. The edge capacitance value $C_\mathrm{edge}$ used in these simulations corresponds to realistic device geometries. We estimate the capacitive coupling between a metallic contact and the hematite edge using a parallel-plate approximation:

\begin{equation}\label{eq:edgeCapacitance}
    C_\mathrm{edge} = \varepsilon_0 \varepsilon_\mathrm{eff} \frac{A}{d},
\end{equation}

where $A = h \cdot l$ is the contact area, with height $h = 300~\mathrm{nm}$ and length $l = 200~\mu\mathrm{m}$, separated from the hematite by a gap $d = 40~\mathrm{nm}$. For a device on a sapphire substrate, the effective dielectric constant is approximately $\varepsilon_\mathrm{eff} \approx \frac{\varepsilon_\mathrm{sapphire}+1}{2} \approx 5.2$, resulting in $C_\mathrm{edge} \approx 70~\mathrm{fF}$. The parasitic capacitance $C_\mathrm{p}$ between the more widely separated feedline arms is correspondingly smaller.

\nobalance 


%

\end{document}